\documentclass[%
 reprint,
 amsmath,amssymb,
 aps,
]{revtex4-2}

\usepackage{bm}

\usepackage{braket}
\usepackage{blindtext}

\usepackage{xcolor}
\usepackage{tikz}
\usepackage[pdftex]{hyperref}
\definecolor{purple}{rgb}{1,0,1}
\definecolor{lime}{HTML}{A6CE39} 

\newcommand{\orcidicon}{%
	\begin{tikzpicture}
	\draw[lime, fill=lime] (0,0) 
		circle [radius=0.16] 
		node[white] {{\fontfamily{qag}\selectfont \tiny ID}};
	\draw[white, fill=white] (-0.0625,0.095) 
		circle [radius=0.007];
	\end{tikzpicture}
	\hspace{-3mm}
}
\newcommand\orcidDel{{\href{https://orcid.org/0000-0003-4158-202X}{\orcidicon}}}

\begin{document}

\preprint{APS/123-QED}

\title{Quantum Analogue of Entropy Based DDoS Detection}

\author{Del Rajan\orcidDel\hspace{-2mm}}

\affiliation{Victoria University of Wellington, \\
Wellington 6140, New Zealand.
\\}

\date{\today}

\begin{abstract}
Distributed Denial-of-Service (DDoS) attacks can occur in quantum networks, which can pose a significant threat to its key distribution protocols.  We introduce a quantum analogue of a classical entropic DDoS detection system, and apply it in the context of detecting an attack on a quantum network.  In particular, we examine DDoS attacks on a quantum repeater and harness the associated entanglement entropy for the detection system.  Our method contributes to the applicability of quantum information from the domain of data security to the area of network security.

\end{abstract}

\maketitle


\textit{Introduction.}---A central aim of quantum information science is to design quantum systems to perform information tasks \cite{NielsenQC}.  Prominent examples of such work involved deriving quantum analogues of classical information applications, and harnessing the quantum resource to demonstrate an advantage over the classical case.  

For instance, the property of superposition is used by quantum models of computation to drastically outperform the best classical supercomputers on certain tasks \cite{arute2019quantum, wu2021strong}.  Another example are the quantum realizations of secure key distribution protocols which are predicated on the impossibility to copy quantum information \cite{xu2020secure}.  Perhaps the most relevant example for our work are the quantum analogues of classical communication networks, simply called quantum networks \cite{wehner2018quantum}.  In these networks, quantum information can be teleported \cite{Bennett1993}, recently to distances exceeding a 1000km \cite{ren2017ground}.

Beyond this established work, a further direction is to design novel quantum technologies by utilizing our understanding of the classical case.  Our main result is to demonstrate progress in this direction.

Distributed Denial-of-Service (DDoS) attacks are a significant topic in classical network security \cite{osterweil202021, mahjabin2017survey}, and various systems have been developed to detect such attacks on a classical network.  Similarly, DDoS attacks have also been identified to be a considerable threat in quantum networks \cite{satoh2021attacking, xu2020secure, robert2021impact}\cite{price2020quantum} and in this work we design a system to detect such attacks.  This advances the applicability of quantum information from the domain of data security to the area of network security.

\textit{Classical DDoS.}---In classical networks, Distributed Denial-of-Service (DDoS) attacks are rather frequent events \cite{osterweil202021, mahjabin2017survey} and their reach can extend into the critical infrastructure of a nation, as witnessed in Estonia \cite{lesk2007new} and New Zealand \cite{nzx}.  In such networks, information is transmitted in the form of data packets and the role of directing this traffic is performed by routers.  To initiate a DDoS flooding attack, many routers would direct packets from multiple attack nodes to a victim node.  The intent behind this flood of traffic is to overload the victim node so that it becomes unresponsive to legitimate traffic.

Preventing a DDoS attack requires most essentially the ability to identify the attack traffic as early as possible \cite{Abigail2019}.  To achieve this temporal capability, detection systems have been designed using the Shannon entropy
\begin{equation}\label{Shannonentropy}
    H(X) \equiv -\sum_{i}p_{i}\log p_{i},
\end{equation}
where $p_{i}$ are the probabilities associated to random variable $X$ and logarithms are taken to base $2$. 

Reviews of such entropic approaches can be found in \cite{mahjabin2017survey, Abigail2019, ozccelik2015deceiving}.  We briefly outline one of these methods \cite{yu2010traceback}.

A flow at a router is group of packets categorized as
\begin{equation}
    f_{ij}(u_{i}, d_{j}, t) \equiv \{ <u_{i}, d_{j},t> | u_{i} \in U, d_{j} \in D\},
\end{equation}
where $i, j \in \mathbb{Z}^+$, $U$ is the set of the upstream routers, $D$ denotes the set of destination addresses from the router, and $t$ is the time stamp.  Let $|f_{ij}(u_{i}, d_{j}, t)|$ represent the number of packets of flow $f_{ij}$ at time $t$.  For a given time interval $\Delta T$, the variation of the number of packets for a given flow is defined as
\begin{equation}
    N_{ij}(u_{i}, d_{j}, t + \Delta t) \equiv |f_{ij}(u_{i}, d_{j}, t + \Delta T)| - |f_{ij}(u_{i}, d_{j}, t)|.
\end{equation}
If $|f_{ij}(u_{i}, d_{j}, t)| = 0$, then $N_{ij}(u_{i}, d_{j}, t + \Delta T)$ is the number of packets of flow $f_{ij}$ that went through the router during time interval $\Delta T$.  The quantity
\begin{equation}
    p_{ij}(u_{i}, d_{j}, t + \Delta T) = \frac{N_{ij}(u_{i}, d_{j}, t + \Delta T)}{\sum_{i=1}^{\infty} \sum_{j=1}^{\infty} N_{ij}(u_{i}, d_{j}, t + \Delta T)},
\end{equation}
gives the probability of the flow $f_{ij}$ over all flows at the router with
\begin{equation}
    \sum_{i=1}^{\infty} \sum_{j=1}^{\infty} p_{ij}(u_{i}, d_{j}, t + \Delta T) = 1.
\end{equation}
The computation of the Shannon entropy (\ref{Shannonentropy}) at the router is obtained through
\begin{equation}\label{EntropyatRouter}
    H(F) = - \sum_{i,j} p_{ij}(u_{i}, d_{j}, t + \Delta T) \log  p_{ij}(u_{i}, d_{j}, t + \Delta T),
\end{equation}
where $F$ is the associated random variable with respect to flows during $\Delta T$.  If the total number of flows is constrained to $N$, then (\ref{EntropyatRouter}) is rather simply
\begin{equation}\label{EntropyatRoutersimple}
    H(F)= H(p_{1}, p_{2}, \ldots, \ p_{N}) = -\sum_{i=1}^{N} p_{i} \log p_{i},
\end{equation}
with $0 \leq H(F) \leq \log N$.  The lower bound occurs when there is only one flow.

In order to model a DDoS attack, a number of assumptions are made.  These are that there is no extraordinary change of traffic in a very short time for the non-attack case, the number of attack packets is at least an order of magnitude higher than that of normal flows, there is only one attack ongoing at a time, and that the number of flows is stable for both non-attack and attack cases.

Suppose attack flows start passing through the router at $t=(n+1)\tau$, hence $t=n\tau$ signifies the time at the router just before the attack.  The respective distributions are
\begin{eqnarray}
 &\{p_{1}^{(n+1)\tau}&, p_{2}^{(n+1)\tau}, \ldots, p_{N}^{(n+1)\tau}\}, \\
&\{p_{1}^{n\tau}&, p_{2}^{n\tau}, \ldots, p_{N}^{n\tau}\}.
\end{eqnarray}
A consequence of the assumptions is that $p_{k}^{n\tau} \ll p_{k}^{(n+1)\tau}$  for some $k$.  Further reasoning with Jensen's inequality leads to
\begin{equation}
    -\sum_{i=1}^{N} p_{i}^{n\tau} \log p_{i}^{n\tau} \gg  -\sum_{i=1}^{N} p_{i}^{(n+1)\tau} \log p_{i}^{(n+1)\tau}.
\end{equation}
Expressing this in terms of the entropy (\ref{EntropyatRoutersimple}) gives 
\begin{equation}\label{classicalDDoSdetection}
    H(F^{{n\tau}}) \gg H(F^{{(n+1)\tau}}).
 \end{equation}
The entropy at the router drops dramatically as soon as attack flows are passing through, thus allowing for an ability to detect an attack as early as possible.

\textit{Quantum Repeaters.}---Quantum networks \cite{wehner2018quantum} generate entanglement over long distances. These entanglement flows are routed through devices known as repeaters \cite{pant2019routing, lee2020quantum} which perform entanglement swapping to connect two spatially entangled links into a longer entangled link.  Alternatively, one can view this procedure as teleporting the entangled quantum information.  

Specifically in this work, we will use Bell states
\begin{equation}\label{spatialquantuminformation}
    \ket{\beta_{xy}} = \frac{\ket{0} \otimes \ket{y} + (-1)^{x}\ket{1} \otimes \ket{\bar{y}}}{\sqrt{2}},
\end{equation}
where the bar denotes negation and we have a choice between $xy = 00, 01, 10,$ or $11$.  With respect to the computational basis states, the quantum information in the Bell state (\ref{spatialquantuminformation}) takes the form
\begin{equation}\label{Inputforinversion}
    \bigg(\frac{\bar{y}}{\sqrt{2}}, \frac{{y}}{\sqrt{2}}, \frac{(-1)^{x}y}{\sqrt{2}}, \frac{(-1)^{x}\bar{y}}{\sqrt{2}}\bigg).
\end{equation}

In our simple routing scenario, the outcome is a Bell state between the request node and the receiver node (which are spatially apart), without the request node directly sending a qubit to the receiver node.  This can be accomplished through a quantum repeater.  We generate Bell pairs at both the request node (qubits $A$ and $B$) and the repeater node (qubits $C$ and $D$).   One qubit ($B$) of the request pair reaches the repeater to be Bell projected with a qubit ($C$) at the repeater.

The joint state can be written as 
\begin{eqnarray}\label{TurnstileStep1}
&\ket{\text{Request}} \otimes \ket{\text{Repeater}} \equiv \ket{\beta_{00}}_{A,B} \otimes \ket{\beta_{xy}}_{C,D}  \nonumber  \\\nonumber
&= \frac{1}{2}(\ket{\beta_{xy}}_{A,D} \otimes \ket{\beta_{00}}_{B,C} \\ 
&+ \ket{\beta_{\bar{x}y}}_{A,D} \otimes \ket{\beta_{10}}_{B,C} \\ \nonumber
&+ (-1)^{x}\ket{\beta_{x\bar{y}}}_{A,D} \otimes \ket{\beta_{01}}_{B,C}\\ \nonumber
&+ (-1)^{x}\ket{\beta_{\bar{x}\bar{y}}}_{A,D} \otimes \ket{\beta_{11}}_{B,C}.
\end{eqnarray}  

The repeater performs a Bell state projection on $BC$.  This returns one of four possible outcomes with consequences for $AD$
\begin{eqnarray}\label{Fourpossibleoutcomes}
\ket{\beta_{00}}_{B,C} \quad &\rightarrow& \quad \ket{\beta_{xy}}_{A,D} \nonumber  \\ 
 \ket{\beta_{01}}_{B,C}\quad &\rightarrow& \quad (-1)^{x}\ket{\beta_{x\bar{y}}}_{A,D} \nonumber \\
 \ket{\beta_{10}}_{B,C}  \quad &\rightarrow& \quad \ket{\beta_{\bar{x}y}}_{A,D} \\ \nonumber
\ket{\beta_{11}}_{B,C}  \quad &\rightarrow& \quad  (-1)^{x}\ket{\beta_{\bar{x}\bar{y}}}. \nonumber
\end{eqnarray}

Depending on the outcome, the repeater applies a particular unitary operator to qubit $D$,
\begin{eqnarray}
&(\mathbb{I} \otimes \mathbb{I}) \ket{\beta_{xy}}_{A,D}, \nonumber \\
&(\mathbb{I} \otimes (-1)^{x}\hat{\sigma}_{1}) (-1)^{x}\ket{\beta_{x\bar{y}}}_{A,D}, \nonumber \\
&(\mathbb{I} \otimes (-1)^{y}\hat{\sigma}_{3})\ket{\beta_{\bar{x}y}}_{A,D},  \\
&(\mathbb{I} \otimes (-1)^{x+y}\hat{\sigma}_{3}\hat{\sigma}_{1}) (-1)^{x}\ket{\beta_{\bar{x}\bar{y}}}_{A,D}, \nonumber
\end{eqnarray}
where $\hat{\sigma}_{1} = \ket{0}\bra{1} + \ket{1}\bra{0}, \hat{\sigma}_{2} = -i\ket{0}\bra{1} + i\ket{1}\bra{0}$ and $\hat{\sigma}_{3} = \ket{0}\bra{0} - \ket{1}\bra{1}$.  Afterwards, the non-projected qubit ($D$) of the repeater pair leaves towards the destination node.  This results in the desired output of having state $\ket{\beta_{xy}}_{A,D}$ shared between the request node and receiver.

We want to view this procedure in terms of the von Neumann quantum entropy \cite{NielsenQC}, which is defined as
\begin{equation}
    S(\rho) \equiv -\text{Tr} (\rho \log \rho),
\end{equation}
where $\rho$ is a density operator.
The entropy is zero if and only if it is a pure state.  For an arbitrary composite system with subsystems $K$ and $L$, the joint and conditional entropy are respectively
\begin{eqnarray}
S(\rho_{KL}) &\equiv -\text{Tr}(\rho_{KL} \log \rho_{LK}), \\
S(\rho_{K}|\rho_{L}) &\equiv S(\rho_{KL}) - S(\rho_{L}),
\end{eqnarray}
where $\rho_{K} = \text{Tr}_{L}(\rho_{KL})$ and $\rho_{L} = \text{Tr}_{K}(\rho_{KL})$.  Suppose $\rho_{KL}$ is a pure state, then $\rho_{KL}$ is entangled if and only if 
\begin{equation}\label{entanglemententropyequivalence}
    S(\rho_{K}|\rho_{L}) < 0.
\end{equation}
In this case the entropy of either subsystem, $S(\rho_{K})$ or $S(\rho_{L})$, is referred to as the entanglement entropy.  

In the repeater, the density operator is
\begin{equation}
  \rho_{C,D}  = \ket{\beta_{xy}}_{B,C} \bra{\beta_{xy}}_{B,C}, 
\end{equation}
and the associated conditional entropy is
\begin{equation}\label{entropyforinput}
    S(\rho_{C}|\rho_{D}) = S(\rho_{C,D}) - S(\rho_{D}) = 0 - 1 = -1,
\end{equation}
which indicates entanglement (\ref{entanglemententropyequivalence}) as expected.  Note that at the time before projection, qubits $A$ and $D$ are not entangled
\begin{equation}
    S(\rho_{A}|\rho_{D}) = S(\rho_{A,D}) - S(\rho_{D}) = 2 - 1 = 1.
\end{equation}
For the output state, the density operator takes the form
\begin{equation}
\rho_{A,D}  = \ket{\beta_{xy}}_{A,D} \bra{\beta_{xy}}_{A,D},    
\end{equation}
and the associated conditional entropy is
\begin{eqnarray}\label{entropyforoutput}
    S(\rho_{A}|\rho_{D}) &=& S(\rho_{A,D}) - S(\rho_{D}) \nonumber \\ 
    &=& 0 - 1 = -1,
\end{eqnarray}
signifying entanglement, with a loss of it in
\begin{eqnarray}\label{inputafterinversion}
    S(\rho_{C}|\rho_{D}) &=& S(\rho_{C,D}) - S(\rho_{D}) \nonumber \\
    &=& 2-1=1.
\end{eqnarray}
In terms of the joint entropy, just before projection (\ref{entropyforinput}) we see that $S(\rho_{C,D}) = 0$ which specifies a pure state.  After projection (\ref{inputafterinversion}) we have that $S(\rho_{C,D}) = 2$ which signifies missing information in $CD$. Part of that missing information moved, as described by the updated entropy $S(\rho_{A,D})= 0$ in the output $\ket{\beta_{xy}}_{A,D}$.

\textit{Quantum DDoS.}---Quantum networks can also experience DDoS attacks \cite{satoh2021attacking},
which poses a significant threat to its quantum key distribution protocols \cite{xu2020secure, robert2021impact}.  Given that repeaters have a maximum session capacity \cite{rabbie2020designing}, we consider DDoS attacks where service requests exceed that maximum capacity. 

The entanglement entropy will be used to formulate a DDoS detection system analogous to the classical case (\ref{classicalDDoSdetection}).  To derive this, we utilize the material from \cite{witten2020mini}.

Our model starts with the request node generating $\ket{\beta_{00}}_{A,B}$ with qubit $B$ is sent to the repeater.  The repeater generates $\ket{\beta_{00}}_{C,D}$ which can be viewed as an instantiation of $\ket{\beta_{xy}}_{C,D}$.  

We consider the quantities before the projection.  The total system is $\rho_{A,B,C,D}$ which denotes
\begin{equation}\label{repeaterinput}
\ket{\beta_{00}}_{A,B} \ket{\beta_{00}}_{C,D}
\bra{\beta_{00}}_{A,B} \bra{\beta_{00}}_{C,D},    
\end{equation}
and the subsystem held at the repeater is $\rho_{B,C,D}$ as it excludes $\rho_{A}$.  Given qubits $C$ and $D$ are jointly in a pure state, we have that
\begin{equation}
S(\rho_{B,C,D}) = S(\rho_{B}) = 1. 
\end{equation}
The qubit $B$ is maximally mixed since it is entangled with qubit $A$ in a Bell state.  Thus the entanglement entropy of qubit $A$ before the projection equates to
\begin{equation}\label{Aentropybeforeprojection}
S(\rho_{A}) = S(\rho_{B,C,D}). 
\end{equation}
We take the partial trace to obtain the density operator for qubit $D$
\begin{equation}
    \rho_{D} = \text{Tr}_{ABC}(\rho_{A,B,C,D}).
\end{equation}
We have that $S(\rho_{D}) = 1$ since it is entangled with $\rho_{C}$.

The entanglement entropy forms a crucial role for a repeater session.  A successful session occurs when the entanglement is swapped (\ref{Fourpossibleoutcomes}). The swapping is successful only if the repeater uses some rank one orthogonal projectors $\Pi_{i}$ such that no matter what outcome occurs at the repeater on qubits $B$ and $C$, the value of the entanglement entropy of $\rho_{A}$ before projection must equal the value of the entanglement entropy of $\rho_{A}$ after projection. 

We proceed to examine the quantities after the projection. If the repeater obtains outcome $i$, then the density operator of $D$ is
\begin{equation}
        \rho_{D_{i}} = \frac{1}{p_{i}}\text{Tr}_{ABC}(\Pi_{i}\rho_{A,B,C,D}),
\end{equation}
where $p_{i}$ is the associated Born probability.  Given
\begin{equation}
    \sum_{i} \Pi_{i} = \mathbb{I},
\end{equation}
we have that
\begin{equation}\label{DentropymadeupofDi}
    \rho_{D} = \sum_{i}p_{i}\rho_{D_{i}}.
\end{equation}
After measurement, qubits $A$ and $D$ are in a pure entangled state. The entanglement entropy of $D$ equates to the entanglement entropy of $A$ after projection  
\begin{equation}\label{Dentropyafterprojection}
    S(\rho_{D_{i}}) = S(\rho_{A}).
\end{equation}
Using the crucial condition that a successful session requires the entropy before projection and entropy after projection of $\rho_{A}$ to equate, we can formulate relationships between the quantities before and after projection.  Specifically using (\ref{Aentropybeforeprojection}) and (\ref{Dentropyafterprojection}), we obtain
\begin{equation}
    S(\rho_{D_{i}}) = S(\rho_{B,C,D}),
\end{equation}
and furthermore
\begin{equation}\label{entropyequality1}
    S(\rho_{B,C,D}) = S(\rho_{D_{i}}) = \sum_{i}p_{i}S(\rho_{D_{i}}).
\end{equation}
Applying the concavity inequality \cite{witten2020mini} to (\ref{DentropymadeupofDi}) results to
\begin{equation}\label{entropyinequality2}
    S(\rho_{D}) \geq \sum_{i}p_{i}S(\rho_{D_{i}}).
\end{equation}
Combining (\ref{entropyequality1}) and (\ref{entropyinequality2}) gives
\begin{equation}
    S(\rho_{B,C,D}) = \sum_{i}p_{i}S(\rho_{D_{i}}) \leq S(\rho_{D}).
\end{equation}
Therefore, before the projection one can predict that a successful session is possible when and only when
\begin{equation}\label{successfulsession}
   S(\rho_{B,C}|\rho_{D}) = S(\rho_{B,C,D}) - S(\rho_{D}) \leq 0.
\end{equation}
A failed session will occur when $S(\rho_{B,C}|\rho_{D}) > 0$.  For our specific case (\ref{repeaterinput}), we have that $S(\rho_{B,C}|\rho_{D}) = 0$ which implies a successful session ahead. 

The capacity of the repeater is defined as the maximum number of sessions it can facilitate simultaneously and this is directly related to the number of Bell pairs that can be stored in memory \cite{rabbie2020designing}.  In our case, this would be the maximum number of copies of $\rho_{C,D}$ generated at a time to keep the service at full capacity.  After that time, any unused Bell pairs get destroyed and the system regenerates to full capacity at the next time point.

With respect to capacity, we modify our previous analysis to $N$ copies of system $\rho_{A,B,C,D}$ for large $N$.  The repeater would make a complete projective measurement on  $(\rho_{B,C})^{\otimes N}$.  In this case all the entropies are multiplied by $N$.  Hence the condition (\ref{successfulsession}) for sessions is maintained and we can interpret it in terms of capacity. 

If $S(\rho_{B,C}|\rho_{D}) < 0$ before the projection, then $-S(\rho_{B,C}|\rho_{D})$ is the number of Bell pairs left afterwards in the memory.  It quantifies the unused capacity after the session requests have been fulfilled. Whereas if $S(\rho_{B,C}|\rho_{D}) > 0$, it not possible to carry out a session to begin with, and predicts an unresponsive service due to requests exceeding the capacity. Therefore (\ref{successfulsession}) can be used to model a DDoS attack.

To design a detection system, suppose a flood of attack requests reaches a repeater at specific time $t=(n+1)\tau$, and as in the classical case \cite{yu2010traceback} we assume the number of attack requests is at least an order of magnitude higher than that of normal requests.  The entropy of the requests at the repeater is quantified as
\begin{equation}\label{quantumattackrequests}
    S(\rho_{B}^{n\tau}) \ll S(\sigma_{B}^{(n+1)\tau}),
\end{equation}
where $\rho$ is used to label the systems involved prior to attack, and $\sigma$ denotes the systems involved in the attack (the superscripts signify the respective times).  The repeater generates the same full capacity at each time point hence
\begin{equation}\label{samecapacity}
 S(\rho_{C,D}^{n\tau, n\tau}) = S(\sigma_{C,D}^{(n+1)\tau, (n+1)\tau}),   
\end{equation}
and with (\ref{quantumattackrequests}) we obtain 
\begin{equation}\label{quantumattackrequests2}
 S(\rho_{B,C,D}^{n\tau, n\tau, n\tau}) \ll S(\sigma_{B,C,D}^{(n+1)\tau, (n+1)\tau, (n+1)\tau}).
 \end{equation}
From (\ref{samecapacity}) we have that 
\begin{equation}
S(\rho_{D}^{n\tau}) = S(\sigma_{D}^{(n+1)\tau}).
\end{equation}
Combining this with (\ref{quantumattackrequests2}) produces an entropic DDoS detection formula at the repeater
 \begin{equation}\label{quantumattackrequests3}
 S(\rho_{B,C}^{n\tau, n\tau}|\rho_{D}^{n\tau}) \ll S(\sigma_{B,C}^{n\tau, n\tau}|\sigma_{D}^{n\tau}).   
\end{equation}  
The conditional entropies in (\ref{quantumattackrequests3}) encode both the requests and the capacity.  In an attack the entropy increases dramatically at the repeater, signifying a drastic reduction in capacity.  This provides an early detection system that is comparable to the classical case (\ref{classicalDDoSdetection}).

\textit{Future Work.}---In classical networks, after detecting a DDoS attack it is common to employ a mitigation method \cite{osterweil202021}. As a result the service capacity is unaffected, leaving service available to legitimate traffic.  For the quantum DDoS case presented, future work could involve developing a mitigation method after detection.

We illustrate this with a simple method that could be used to develop a more sophisticated strategy.  Suppose we have attack Bell states  $\ket{\beta_{00}}_{A_{1}B_{1}}^{(n+1)\tau, (n+1)\tau}$ and $\ket{\beta_{00}}_{A_{2}B_{2}}^{(n+1)\tau, (n+1)\tau}$.  The labels $A_{1}$ and $A_{2}$ refer to the qubits at the respective attack nodes, and labels $B_{1}$ and $B_{2}$ are the qubits that reach the repeater at $t=(n+1)\tau$.  Under normal conditions, the repeater would perform a projection with the Bell pairs generated at the repeater.  

Given that the attack has been detected, the repeater rather performs a joint projective measurement on the attack qubits themselves, $B_{1}$ and $B_{2}$ at $t=(n+1)\tau$.  We can write this as 
\begin{eqnarray}
&\ket{\beta_{00}}_{A_{1}B_{1}}^{(n+1)\tau, (n+1)\tau} \otimes  \ket{\beta_{00}}_{A_{2}B_{2}}^{(n+1)\tau, (n+1)\tau}  \nonumber  \\
&=  \frac{1}{2}( \ket{\beta_{00}}_{A_{1}A_{2}}^{(n+1)\tau, (n+1)\tau} \otimes \ket{\beta_{00}}_{B_{1}B_{2}}^{(n+1)\tau, (n+1)\tau} \\ 
&+ \ket{\beta_{01}}_{A_{1}A_{2}}^{(n+1)\tau, (n+1)\tau} \otimes \ket{\beta_{01}}_{B_{1}B_{2}}^{(n+1)\tau, (n+1)\tau} \\ \nonumber
&+ \ket{\beta_{10}}_{A_{1}A_{2}}^{(n+1)\tau, (n+1)\tau} \otimes \ket{\beta_{10}}_{B_{1}B_{2}}^{(n+1)\tau, (n+1)\tau}\\ \nonumber
&+ \ket{\beta_{11}}_{A_{1}A_{2}}^{(n+1)\tau, (n+1)\tau} \otimes \ket{\beta_{11}}_{B_{1}B_{2}}^{(n+1)\tau, (n+1)\tau}.
\end{eqnarray}  
This joint measurement would swap the entanglement to qubits $A_{1}$ and $A_{2}$ which are at the attack nodes.  Consequently the Bell pairs generated at the repeater have not been used, thereby leaving the repeater capacity available for legitimate traffic. 

In a large classical network, it advantageous to devise a traceback method \cite{yu2010traceback} so to identify the source of the attack.  Both classical and quantum networks can be modelled as a directed acyclic graphs, where the upstream routers could be viewed as parent nodes and the downstream routes as children nodes.  Hence an interesting direction would be to employ quantum causal models \cite{barrett2019quantum} to provide a traceback model for DDoS attacks on a quantum network.  The attacks could be formulated in terms of do-interventions, and would allow the ability to apply a quantum do-calculus or a quantum causal discovery algorithm to carry out successful traceback for a DDoS attack on a quantum network.

\textit{Conclusion.}---DDoS attacks are a central topic in classical network security, and have been identified to be significant threat on quantum networks.  In this work, we design a quantum analogue of a classical DDoS detection system, and apply it in the context of a quantum network.  We hope that our  design contributes to extending the applicability of quantum information from the domain of data security to area of network security.



\end{document}


\preprint{APS/123-QED}

\title{Supplemental Material to: Quantum Analogue of Entropy Based DDoS Detection}

\section{DERIVATION OF THE QUANTUM REPEATER EQUATIONS}

In main text, we had the equation

\begin{eqnarray}
 \ket{\beta_{00}}_{A,B} \otimes \ket{\beta_{xy}}_{C,D}
 &=& \frac{1}{2}(\ket{\beta_{xy}}_{A,D} \otimes \ket{\beta_{00}}_{B,C}  
+ \ket{\beta_{\bar{x}y}}_{A,D} \otimes \ket{\beta_{10}}_{B,C} \\ \nonumber
&+& (-1)^{x}\ket{\beta_{x\bar{y}}}_{A,D} \otimes \ket{\beta_{01}}_{B,C} 
+ (-1)^{x}\ket{\beta_{\bar{x}\bar{y}}}_{A,D} \otimes \ket{\beta_{11}}_{B,C}.
\end{eqnarray}

It can be derived in the following manner:
\begin{eqnarray}
 \ket{\beta_{00}}_{A,B}  \ket{\beta_{xy}}_{C,D}
 &=& \Bigg(\frac{\ket{00}_{A,B} + \ket{11}_{A,B}}{\sqrt{2}}\Bigg) \otimes
 \Bigg(\frac{\ket{0y}_{C,D} + (-1)^{x} \ket{1\bar{y}}_{C,D}}{\sqrt{2}}\Bigg)  \\
 &=& \frac{1}{2}(\ket{00}_{A,B}\ket{0y}_{C,D}
 + (-1)^{x}\ket{00}_{A,B} \ket{1\bar{y}}_{C,D} \\
 &+& \ket{11}_{A,B}\ket{0y}_{C,D}
 + (-1)^{x}\ket{11}_{A,B} \ket{1\bar{y}}_{C,D}) \\
 &=& \frac{1}{2}(\ket{0y}_{A,D}\ket{00}_{B,C}
 + (-1)^{x}\ket{0\bar{y}}_{A,D} \ket{01}_{B,C} \\
 &+& \ket{1y}_{A,D}\ket{10}_{B,C}
 + (-1)^{x}\ket{1\bar{y}}_{A,D} \ket{11}_{B,C}) \\
 &=& \frac{1}{4}\bigg[2\ket{0y}_{A,D}\ket{00}_{B,C} +
 2(-1)^{x}\ket{1\bar{y}}_{A,D} \ket{11}_{B,C} \\
 &+& 2(-1)^{x}\ket{0\bar{y}}_{A,D} \ket{01}_{B,C}
 + 2\ket{1y}_{A,D}\ket{10}_{B,C}\bigg] \\
 &=& \frac{1}{2} \bigg[ \frac{1}{2}(\ket{0y}_{A,D}\ket{00}_{B,C}
 + \ket{0y}_{A,D} \ket{11}_{B,C} \\
 &+& (-1)^{x}\ket{1\bar{y}}_{A,D}\ket{00}_{B,C}
 + (-1)^{x}\ket{1\bar{y}}_{A,D} \ket{11}_{B,C})\\
 &+& \frac{1}{2}(\ket{0y}_{A,D}\ket{00}_{B,C}
 - \ket{0y}_{A,D} \ket{11}_{B,C}\\
 &+& (-1)^{\bar{x}}\ket{1\bar{y}}_{A,D}\ket{00}_{B,C}
 - (-1)^{\bar{x}}\ket{1\bar{y}}_{A,D} \ket{11}_{B,C}) \\
 &+& \frac{1}{2}((-1)^{x}\ket{0\bar{y}}_{A,D}\ket{01}_{B,C}
 + (-1)^{x}\ket{0\bar{y}}_{A,D} \ket{10}_{B,C} \\
 &+& \ket{1y}_{A,D}\ket{01}_{B,C}
 + \ket{1y}_{A,D} \ket{10}_{B,C}) \\
 &+& \frac{1}{2}((-1)^{x}\ket{0\bar{y}}_{A,D}\ket{01}_{B,C}
 - (-1)^{x}\ket{0\bar{y}}_{A,D} \ket{10}_{B,C} \\
 &-& \ket{1y}_{A,D}\ket{01}_{B,C}
 + \ket{1y}_{A,D} \ket{10}_{B,C}) \bigg] \\
 &=& \frac{1}{2} \Bigg[\Bigg(\frac{\ket{0y}_{A,D}+ (-1)^{x}\ket{1\bar{y}}_{A,D}}{\sqrt{2}}\Bigg)
 \Bigg(\frac{\ket{00}_{B,C}+ \ket{11}_{B,C}}{\sqrt{2}}\Bigg) \\
 &+& \Bigg(\frac{\ket{0y}_{A,D} + (-1)^{\bar{x}}\ket{1\bar{y}}_{A,D}}{\sqrt{2}}\Bigg)
 \Bigg(\frac{\ket{00}_{B,C} -  \ket{11}_{B,C}}{\sqrt{2}}\Bigg) \\
  &+& \Bigg(\frac{(-1)^{x}\ket{0\bar{y}}_{A,D}+ (-1)^{2x}\ket{1y}_{A,D}}{\sqrt{2}}\Bigg)
 \Bigg(\frac{\ket{01}_{B,C} +  \ket{10}_{B,C}}{\sqrt{2}}\Bigg) \\
&+& \Bigg(\frac{(-1)^{x}\ket{0\bar{y}}_{A,D} + (-1)^{x + \bar{x}}\ket{1y}_{A,D}}{\sqrt{2}}\Bigg)
 \Bigg(\frac{\ket{01}_{B,C} - \ket{10}_{B,C}}{\sqrt{2}}\Bigg)
 \Bigg] \\
 &=& \frac{1}{2}(\ket{\beta_{xy}}_{A,D} \otimes \ket{\beta_{00}}_{B,C}
+ \ket{\beta_{\bar{x}y}}_{A,D} \otimes \ket{\beta_{10}}_{B,C} \\ \nonumber
&+& (-1)^{x}\ket{\beta_{x\bar{y}}}_{A,D} \otimes \ket{\beta_{01}}_{B,C}
+ (-1)^{x}\ket{\beta_{\bar{x}\bar{y}}}_{A,D} \otimes \ket{\beta_{11}}_{B,C}.
\end{eqnarray}

We also applied the following operators to the respective states to obtain the outcome $\ket{\beta_{xy}}_{A,D}$:

\begin{eqnarray}
(\mathbb{I} \otimes (-1)^{x}\hat{\sigma}_{1}) (-1)^{x}\ket{\beta_{x\bar{y}}}_{A,D} &=& (\mathbb{I} \otimes (-1)^{x}\hat{\sigma}_{1})  \Bigg(\frac{(-1)^{x}\ket{0\bar{y}}_{A,D} - \ket{1y}_{A,D}}{\sqrt{2}}\Bigg)  \\
&=& \frac{1}{\sqrt{2}}(\ket{0}_{A} \otimes (-1)^{x}(-1)^{x}\hat{\sigma}_{1}\ket{\bar{y}}_{D}) \\ 
&+& \frac{1}{\sqrt{2}}(\ket{1}_{A} \otimes (-1)^{x}\hat{\sigma}_{1}\ket{y}_{D}) \\
&=& \frac{1}{\sqrt{2}}(\ket{0}_{A} \otimes (-1)^{x+x}\ket{y}_{D}) \\
&+&\frac{1}{\sqrt{2}}(\ket{1}_{A} \otimes (-1)^{x}\ket{\bar{y}}_{D}) \\
&=& \frac{1}{\sqrt{2}}(\ket{0}_{A} \otimes \ket{y}_{D}) \\
&+&\frac{1}{\sqrt{2}}((-1)^{x}\ket{1}_{A} \otimes \ket{\bar{y}}_{D}) \\
&=&\Bigg(\frac{\ket{0y}_{A,D} + (-1)^{x} \ket{1\bar{y}}_{A,D}}{\sqrt{2}}\Bigg) \\
&=& \ket{\beta_{xy}}_{A,D}
\end{eqnarray}

\begin{eqnarray}
(\mathbb{I} \otimes (-1)^{y}\hat{\sigma}_{3})\ket{\beta_{\bar{x}y}}_{A,D} &=& (\mathbb{I} \otimes (-1)^{y}\hat{\sigma}_{3})  \Bigg(\frac{\ket{0y}_{A,D} +(-1)^{\bar{x}}\ket{1\bar{y}}_{A,D}}{\sqrt{2}}\Bigg)\\
&=& \frac{1}{\sqrt{2}}(\ket{0}_{A} \otimes (-1)^{y}\hat{\sigma}_{3}\ket{y}_{D}) \\ 
&+& \frac{1}{\sqrt{2}}(\ket{1}_{A} \otimes (-1)^{y}(-1)^{\bar{x}}\hat{\sigma}_{3}\ket{\bar{y}}_{D}) \\
&=& \frac{1}{\sqrt{2}}(\ket{0}_{A} \otimes (-1)^{y}(-1)^{y}\ket{y}_{D}) \\ 
&+& \frac{1}{\sqrt{2}}(\ket{1}_{A} \otimes (-1)^{y}(-1)^{\bar{x}}(-1)^{\bar{y}}\ket{\bar{y}}_{D}) \\
&=& \frac{1}{\sqrt{2}}(\ket{0}_{A} \otimes (-1)^{y+y}\ket{y}_{D}) \\ 
&+& \frac{1}{\sqrt{2}}(\ket{1}_{A} \otimes (-1)^{y+\bar{y}}(-1)^{\bar{x}}\ket{\bar{y}}_{D}) \\
&=& \frac{1}{\sqrt{2}}(\ket{0}_{A} \otimes \ket{y}_{D}) \\ 
&+& \frac{1}{\sqrt{2}}(\ket{1}_{A} \otimes (-1)(-1)^{\bar{x}}\ket{\bar{y}}_{D}) \\
&=& \frac{1}{\sqrt{2}}(\ket{0}_{A} \otimes \ket{y}_{D}) \\ 
&+& \frac{1}{\sqrt{2}}(\ket{1}_{A} \otimes (-1)^{x}\ket{\bar{y}}_{D}) \\
&=&\Bigg(\frac{\ket{0y}_{A,D} + (-1)^{x} \ket{1\bar{y}}_{A,D}}{\sqrt{2}}\Bigg) \\
&=& \ket{\beta_{xy}}_{A,D}
\end{eqnarray}

\begin{eqnarray}
(\mathbb{I} \otimes (-1)^{x+y}\hat{\sigma}_{3}\hat{\sigma}_{1}) (-1)^{x}\ket{\beta_{\bar{x}\bar{y}}}_{A,D} &=& (\mathbb{I} \otimes (-1)^{x+y}\hat{\sigma}_{3}\hat{\sigma}_{1}) \Bigg(\frac{(-1)^{x}\ket{0\bar{y}}_{A,D} -\ket{1y}_{A,D}}{\sqrt{2}}\Bigg)\\
&=& \frac{1}{\sqrt{2}}(\ket{0}_{A} \otimes (-1)^{x}(-1)^{x+y}\hat{\sigma}_{3}\hat{\sigma}_{1}\ket{\bar{y}}_{D}) \\ 
&-& \ket{1}_{A} \otimes (-1)^{x+y}\hat{\sigma}_{3}\hat{\sigma}_{1}\ket{y}_{D}) \\
&=& \frac{1}{\sqrt{2}}(\ket{0}_{A} \otimes (-1)^{x}(-1)^{x+y}(-1)^{y}\ket{y}_{D}) \\ 
&-& \ket{1}_{A} \otimes (-1)^{x+y}(-1)^{\bar{y}}\ket{\bar{y}}_{D}) \\
&=& \frac{1}{\sqrt{2}}(\ket{0}_{A} \otimes (-1)^{x+y + x +y}\ket{y}_{D}) \\ 
&-& \ket{1}_{A} \otimes (-1)^{x+y+\bar{y}}\ket{\bar{y}}_{D}) \\
&=& \frac{1}{\sqrt{2}}(\ket{0}_{A} \otimes \ket{y}_{D}) \\ 
&-& \ket{1}_{A} \otimes -(-1)^{x}\ket{\bar{y}}_{D}) \\
&=& \frac{1}{\sqrt{2}}(\ket{0}_{A} \otimes \ket{y}_{D}) \\ 
&+& (-1)^{x} \ket{1}_{A} \otimes \ket{\bar{y}}_{D}) \\
&=&\Bigg(\frac{\ket{0y}_{A,D} + (-1)^{x} \ket{1\bar{y}}_{A,D}}{\sqrt{2}}\Bigg) \\
&=& \ket{\beta_{xy}}_{A,D}
\end{eqnarray}